\documentclass[12pt]{article}
\usepackage{amsfonts,amsthm,amsmath,amssymb,upgreek,bm}
\usepackage{hyperref}
\usepackage[paper=letterpaper,margin=.85in]{geometry}
\usepackage{graphicx}
\usepackage{units}
\usepackage{float}
\usepackage[export]{adjustbox}
\usepackage[dvipsnames]{xcolor}
\usepackage[shortlabels]{enumitem}
\usepackage[mathscr]{euscript}
\usepackage[normalem]{ulem}
\usepackage{bm}
\usepackage{tikz}
\usepackage{tikzsymbols}
\usepackage{pgfplots}
\usepackage{ytableau}
\usepackage{slashed}
\usepackage{mathtools}
\usepackage{bbold}
\usetikzlibrary{decorations.pathmorphing}
\usetikzlibrary{arrows.meta}
\usetikzlibrary{calc}
\usepgfplotslibrary{fillbetween}
\usepackage[nameinlink]{cleveref}
\pgfplotsset{compat=1.18}

\newcommand{\be}{\begin{equation}}
\newcommand{\ee}{\end{equation}}
\newcommand{\bea}{\begin{eqnarray}}
\newcommand{\eea}{\end{eqnarray}}
\renewcommand{\tilde}{\widetilde}
\renewcommand{\i}{\mathrm{i}}
\renewcommand{\d}{\mathrm{d}}

\numberwithin{equation}{section}

\begin{document}
\thispagestyle{empty}

\vspace*{2.5cm}
\begin{center}

{\bf {\LARGE On the phase of the de Sitter density of states}}

\begin{center}

\vspace{1cm}

{Yiming Chen${}^1$, Douglas Stanford${}^1$, Haifeng Tang${}^1$, and Zhenbin Yang${}^{2}$}\\
 \bigskip \rm

\bigskip 

{${}^1$  Leinweber Institute for Theoretical Physics, Stanford University, Stanford, CA 94305, USA}
\,\vspace{0.5cm}\\
{${}^2$ Institute for Advanced Study, Tsinghua University, Beijing 100084, China}

\rm
  \end{center}

\vspace{2.5cm}
{\bf Abstract}
\end{center}
\begin{quotation}
\noindent

The one-loop gravitational path integral around Euclidean de Sitter space $S^D$ has a complex phase that casts doubt on a state counting interpretation. Recently, it was proposed to cancel this phase by including an observer. We explore this proposal in the case where the observer is a charged black hole in equilibrium with the de Sitter horizon. We compute the phase of the one-loop determinant within a two-dimensional dilaton gravity reduction, using both numerical and analytical methods. Our results interpolate between previous studies of a probe geodesic observer and the Nariai solution. We also revisit the prescription for going from the Euclidean path integral to the state-counting partition function, finding a positive sign in the final density of states.

\end{quotation}

\setcounter{page}{0}
\setcounter{tocdepth}{2}
\setcounter{footnote}{0}

\newpage

\setcounter{page}{2}
\tableofcontents

\newpage

\section{Introduction}
The round sphere $S^D$ is a simple solution of Euclidean gravity with a positive cosmological constant. The on-shell action \cite{Gibbons:1976ue} equals minus the de Sitter entropy \cite{Gibbons:1977mu}, so it is tempting to regard the sphere as a partition function counting the number of states in some fine-grained description of de Sitter space. 

A challenge to this idea comes from Polchinski's computation of the one-loop determinant, which was found to be proportional to $\i^{D+2}$ \cite{Polchinski:1988ua}, rather than being real and positive as expected for a count of states. This phase arises from a combination of two facts. First, the conformal modes of the metric are  mostly negative modes that should be Wick-rotated in order to make the path integral convergent \cite{Gibbons:1978ac}. Naively, this leads to a factor of $(-\i)^\infty$ that can be absorbed into the definition of an ultralocal measure. Second, $(D+2)$ of the conformal modes actually do {\it not} need to be Wick-rotated. These are the $\ell = 0$ and $\ell = 1$ modes of the conformal factor. The $\ell = 0$ mode represents a physical but stable mode -- the overall size of the sphere. The $\ell = 1$ modes are pure gauge -- they correspond to the $D+1$ conformal Killing vectors of the sphere that are not Killing vectors.\footnote{In de Donder gauge, one adds a gauge fixing term that turns these into positive modes. In the approach of \cite{Law:2020cpj,Law:2025yec}, one simply doesn't integrate over these modes. Either way, they do not require Wick-rotation.} All together, one finds $(-\i)^{\infty - 1 - (D+1)} \to \i^{D+2}$.

This puzzle was addressed by Maldacena in \cite{Maldacena:2024spf}, motivated in part by the usefulness of including an observer in other studies of de Sitter space \cite{Anninos:2011af,Anninos:2017hhn,Chandrasekaran:2022cip,Witten:2023xze}.\footnote{It would be interesting to connect this proposal to the bulk vs.~edge mode discussion in \cite{Anninos:2020hfj,Law:2025ktz}.} The idea of \cite{Maldacena:2024spf} is that if an observer is included, all but two of the $D+1$ conformal Killing vectors become physical modes, because they move the observer. Or, in a different gauge, the observer adds $D-1$ new unstable modes that require Wick rotation. Either way, the phase now becomes $\i^{D+2 - (D-1)} = \i^{3} = -\i$.\footnote{Further work studied the phase for product manifolds \cite{Shi:2025amq,Ivo:2025yek} and the Coleman-de Luccia instanton \cite{Ivo:2025fwe}.}

The answer $-\i$ is still not appropriate for the phase of a sum over states. A final factor of $\i$ was also explained in \cite{Maldacena:2024spf}, following \cite{Casali:2021ewu,Marolf:2022ybi} (see also \cite{PhysRevD.18.2733}). Essentially, this is the same factor of $\i$ that has to be included in the inverse Laplace transform integral over $\beta$ that computes the microcanonical density of states for any thermal system. In the de Sitter case, this integral over $\beta$ imposes the constraint $H = 0$. A wrinkle is that in \cite{Maldacena:2024spf} the final answer for the state-counting partition function came out real but negative.

Our paper has two main points. {\bf First}, we follow up on a suggestion in \cite{Maldacena:2024spf} to replace the observer by a charged black hole in thermal equilibrium with the de Sitter horizon. 
\begin{itemize}
\item In \cref{sec:chargedDeSitterBlackHoles} we review the phase diagram of four-dimensional magnetically charged black holes in de Sitter space \cite{Mellor:1989gi,Romans:1991nq,Mann:1995vb,Booth:1998gf,Montero:2019ekk,Morvan:2022aon}, and their dilaton gravity reductions \cite{Castro:2022cuo}.
\item In \cref{sec:phasePF} we numerically compute the phase of the one-loop partition functions of these dilaton gravity theories. We expect but do not prove that this agrees with the phase of the full higher dimensional one-loop determinant \cite{Haifeng}. For the lukewarm and stable Nariai black holes we get the same answer $Z/|Z| = -\i$ that one finds for de Sitter with an observer, as anticipated in \cite{Maldacena:2024spf}. For the unstable Nariai branch, $Z$ is real and positive, matching onto \cite{Shi:2025amq,Ivo:2025yek,Law:2025yec}. The three branches meet at a point where the one-loop determinant diverges. We also analytically study the ultracold limit, where gravitational fluctuations become large, obtaining the same answer for the phase.
\end{itemize}
{\bf Second}, we address the negativity of the state-counting partition function.
\begin{itemize}
    \item In \cref{sec:statecount} we argue that the idea of \cite{Maldacena:2024spf} actually leads to a real and positive state-counting partition function. The difference is that we do the integral over $\beta$ before a final energy integral, following \cite{Marolf:2022ybi}. The integrals are conditionally convergent with this order of integration, and the result is positive.
\end{itemize}
In the end, we believe that the idea of \cite{Maldacena:2024spf} works as originally intended, and that the density of states in de Sitter is positive once an observer is included.

\section{Charged de Sitter black holes}\label{sec:chargedDeSitterBlackHoles}
A black hole can provide a convenient model of an observer. For one thing, it is native to the gravitational theory, without the need to add new degrees of freedom. For another thing, a black hole automatically carries a clock that describes its age and has finite entropy. Finally, in de Sitter space, a charged black hole can be in a stable thermal equilibrium with the cosmic horizon. In this section, we will review the phase diagram of charged black holes in dS and discuss their dilaton gravity reductions.

\begin{figure}
    \centering
    \includegraphics[width=0.55\linewidth, valign = c]{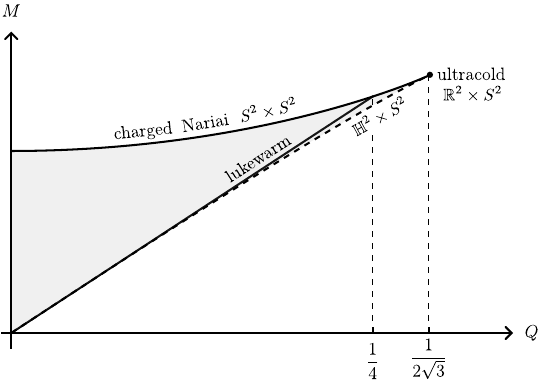}
    \hspace{20pt}
    \begin{tabular}{|c|c|c|c|}
     \hline
     type &  $Z/|Z|$\\
     \hline\hline
     lukewarm &  $-\i$\\
     \hline
     Nariai $Q<1/4$ &   $1$\\
     \hline
     Nariai $Q > 1/4$ &  $-\i$\\
     \hline
    \end{tabular}
    \caption{{\bf Left:} The shark fin diagram of de Sitter Reissner Nordstr\"{o}m black holes. Physical black hole solutions only exist in the shaded region of the $M-Q$ plane. Equilibrium black holes with smooth Euclidean geometries are along the lukewarm line and the Nariai curve. Along the dashed curve are solutions $\mathbb{H}^2\times S^2$. {\bf Right:} Our results, computed below, for the phase of the corresponding dilaton gravity partition functions.}
    \label{fig:sharkfin}
\end{figure}

In four-dimensional Einstein-Maxwell theory, one can consider either electrically or magnetically charged Reissner-Nordstr\"{o}m black holes  \cite{Romans:1991nq,Mann:1995vb,Booth:1998gf}. The  action has the form
\begin{equation}
    I_{\rm 4D} =  \frac{1}{16 \pi G_N}\int \d^4 x\, \sqrt{-g} \left(R - 2 \Lambda - F^2\right)\,.
\end{equation}
where $\Lambda = 3/\ell_{\text{dS}}^2$ and we will set $\ell_\text{dS} = 1$ in the following. The electric case was discussed recently in \cite{Castro:2022cuo}. We will focus on the magnetic case because the continuation to Euclidean signature is simpler. The solution is characterized by two parameters $M$ and $Q$:\footnote{$Q$ is related to the integer-quantized charge $\mathbf{q}$ by $Q^2 = \pi G \mathbf{q}^2$.}
\begin{equation}\label{solns}
\begin{aligned}
    \d s^2  & = -f(r) \d t^2 + \frac{\d r^2}{f(r)} + r^2 ( \d \theta^2 + \sin^2 \theta \d\varphi^2 )\,, \quad \quad F   = Q \sin\theta\, \d\theta \wedge \d \varphi\, ,\\
  f(r) & = 1 - \frac{2M}{r} + \frac{Q^2}{r^2} - r^2\,  .
\end{aligned}
\end{equation}
The physical parameter space of such solutions is the gray ``shark fin'' region shown in \cref{fig:sharkfin}. The black hole is in thermal equilibrium with the cosmic horizon along the lukewarm line where $M = Q$, as well as the charged Nariai line. The two loci intersect at $Q=\frac{1}{4}$. Along these lines, we get a smooth Euclidean solution where the time circle shrinks smoothly at both the cosmic horizon and the outer horizon of the black hole. These equilibria are stable everywhere along the lukewarm line and along the portion of the Nariai curve with $Q > \frac{1}{4}$. The Nariai curve with $Q< \frac{1}{4}$ is unstable and will evaporate towards the lukewarm line \cite{Bousso:1999ms,Montero:2019ekk}. The diagram also contains a (dashed) line which sets the lower limit of the mass for given charge, where the black hole becomes extremal while the cosmic horizon remains at finite temperature. The Nariai line joins the extremal black hole line at $Q = \frac{1}{2\sqrt{3}}$, in the so-called ultra-cold limit. We will make more comments on this limit in  \cref{sec:ultracold}. 

\begin{figure}
    \centering
    \includegraphics[width=0.5\linewidth]{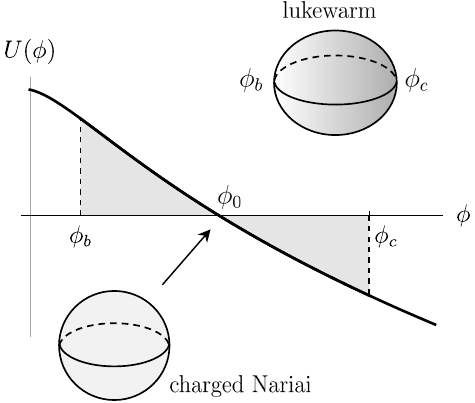}
    \caption{For a dilaton potential $U(\phi)$ with a given $Q$, there are two solutions we can consider. On the upper right corner, we have the lukewarm solution where the dilaton $\phi$ varies from $\phi_b$ at the black hole horizon to $\phi_c$ at the cosmic horizon, where $\phi_b, \phi_c$ satisfy $\int_{\phi_b}^{\phi_c} U  = 0$ and $U(\phi_c) =- U(\phi_b)$. We also have the charged Nariai solution  with a constant dilaton $\phi_0$ at which the potential vanishes, i.e., $U(\phi_0) =0$.}
    \label{fig:potential}
\end{figure}

Since the solutions are spherically symmetric, they admit a dimensional reduction to solutions to a two-dimensional dilaton gravity. We will set the gauge field to its on-shell value in (\ref{solns}) before performing the dimensional reduction. In other words, we are not keeping the gauge field as a dynamical degree of freedom in the 2D theory. Writing the 4D metric as 
\begin{equation}\label{4d2d}
    \d s_{\rm 4D}^2 = \phi^{-\frac{1}{2 }} \d s_{\rm 2D}^2 + \phi \d\Omega_2^2
\end{equation}
and performing dimensional reduction, we arrive at the two-dimensional action
\begin{equation}\label{eqn:dilatonac}
    I = -\frac{1}{4G}\int \d^2 x \sqrt{g}(\phi R + U(\phi))\,,
\end{equation}
with a dilaton potential $U(\phi)$ that depends on the charge of the black hole
\begin{equation}\label{Uphi}
U(\phi) =    \frac{2}{\sqrt{\phi} } \left(  1  - 3 \phi  - \frac{Q^2}{\phi} \right)\,. 
\end{equation}
Solutions in general dilaton gravity theories can be written as\footnote{Note that $\tau$ in the 2D metric is related to $t$ in the 4D metric (\ref{solns}) by $2\tau = \i t$.} 
\be
g_{\mu\nu} \d x^\mu \d x^\nu = \frac{\d\phi^2}{A(\phi)} + A(\phi)\d\tau^2, \hspace{20pt} A'(\phi) = U(\phi), \hspace{20pt} \tau\sim\tau + \beta\,.
\label{eq:  background metric}
\ee
We can get the explicit form of $A(\phi)$ by integrating (\ref{Uphi}),
\be\label{soln2d}
A(\phi) = \frac{4}{\sqrt{\phi}}\left(\phi-\phi^2+Q^2\right) - 8 M\,,
\ee
where $M$ is an integration constant that is the same as the mass of the black hole in 4D. One can check that, with $\phi = r^2$ and the ansatz (\ref{4d2d}), one indeed recovers the metric in (\ref{solns}). For the charged Nariai solution, where the dilaton is a constant, we need to choose a different coordinate (rather than $\phi$) to parameterize the solution. We illustrate both the lukewarm and the charged Nariai solutions in~\cref{fig:potential}.

Our goal will be to understand the phase of the one-loop determinant for the solutions in various parts of the shark fin diagram, within the truncated dilaton model. Of course, there is a priori no guarantee that the phase of the one-loop determinant in the truncated model will agree with the full higher dimensional analysis. However, as we will see, the answer from the 2D theory does agree with known higher dimensional results in some special limits, such as for the case of uncharged Nariai black holes \cite{Ivo:2025yek}. We believe that this is not an accident, but rather that the two-dimensional truncation already captures all the ``problematic" modes that require Wick rotation (after already Wick rotating a whole field).\footnote{Note that this is \emph{not} true for pure de Sitter space, since there are conformal Killing vector modes that contribute to the phase that are not in the s-wave sector. However, these are the precisely the conformal Killing vector modes that are lifted (or canceled) by the introduction of an observer or a black hole \cite{Maldacena:2024spf}.} We also don't expect additional physical instabilities associated with the gauge field or other Kaluza-Klein modes that would lead to additional phases. Nonetheless, it would be more satisfying to have a complete higher dimensional analysis, which is under progress in \cite{Haifeng}.

{\bf Note:} we learned that some related questions regarding heat capacities, EFT breakdown and charged decay close to the classical Nariai and lukewarm states are being considered in \cite{Jan}.

\section{The phase of the Euclidean path integral}\label{sec:phasePF}

\subsection{Setting up the computation}\label{sec:setup}

In this section, we study the quadratic fluctuations of dilaton gravity expanded around a lukewarm or Nariai solution. We will work in the conformal gauge, where 
\begin{equation}\label{fluc}
     \phi = \hat{\phi} + \varphi, \quad g = e^{2\omega} \hat{g}
\end{equation} 
and $\hat{\phi}, \hat{g}$ denote the classical solution while $\varphi, \omega$ represent the fluctuations. The quadratic action is 
\begin{equation}\label{quad}
\begin{aligned}
    I & =  \frac{1}{4G} \int \d^2 x \sqrt{\hat{g}}\, ( \varphi  , \,\, \omega )  \mathcal{D} \begin{pmatrix} \varphi  \\  \omega \end{pmatrix}   , \quad \textrm{where} \quad   \mathcal{D} & = \begin{pmatrix}  - U''/2   &   \hat{\nabla}^2  -  U' \\
     \hat{\nabla}^2    -  U' & -2U  \end{pmatrix}  .
\end{aligned}
\end{equation}
In going to the conformal gauge, we also pick up a ghost determinant, which does not contribute to the phase. It is easy to spot the conformal factor problem in (\ref{quad}). When we go to the UV limit, where the action is dominated by the kinetic term, we could go to a basis $\omega \pm \varphi$ where the action becomes diagonal. The $\omega + \varphi$ field becomes negative while the $\omega - \varphi$ field becomes positive. Away from the UV, the $\omega$ and $\varphi$ fields are coupled so one cannot easily identify a field that is negative.

In the one-loop determinant, each negative mode gives a factor of $(-\i)$, see \cref{app:negative}. Our starting point in analyzing the phase is to define the measure so that it is real and positive if we Wick-rotate the entire conformal factor, or more generally one whole field \cite{Gibbons:1978ac}. In other words, we multiply the naive one-loop determinant by $(+\i)^{n}$, where $n$ is the number of spacetime points. To evaluate the net phase, we would like to compare the number of negative eigenvalues of $\mathcal{D}$ to the number of spacetime points. In \cref{sec:mesh}, we use a pragmatic approach\footnote{One could instead simply rotate one or the other of $\phi,\omega$ (or a linear combination) of them, but this leads to a complex fluctuation operator $\mathcal{D}$ that we found difficult to analyze. We also tried to define the regularized number of negative eigenvalues using a smooth cutoff in eigenvalue space, e.g.~$\sum_{\lambda < 0} e^{-\epsilon^2\lambda^2}$ and then dropping the $1/\epsilon$ term. However, because there are finite counterterms, this is not particularly unique, and in fact one finds that the answer depends on $Q$.} that allows this comparison to be made. We discretize the manifold with a mesh, on which we study the spectrum of the quadratic fluctuations. For a finite mesh, the number of negative eigenvalues of $\mathcal{D}$ and the number of spacetime points are both finite and can be compared. We check this numerical answer with an analytical computation in the ultracold limit and an analytical understanding of how the phase changes at the special point $Q = 1/4$.

There are certain features of the spectrum, specifically, the number of zero modes, which we can anticipate without doing any computation. Apart from possible physical zero modes, we would have additional zero modes whenever the conformal gauge does not remove the gauge redundancies completely, so there are residual gauge transformations in the fluctuations $\varphi, \omega$. 

In fact, this always happens when we are considering the sphere topology. On a two-sphere topology, there are six conformal Killing vectors (CKV) which generate diffeomorphisms that are not fixed by the conformal gauge, namely, the transformed metric has the form $g = e^{2\omega_{\rm{CKV}}} \hat{g}$ so is allowed under the gauge (\ref{fluc}). To properly define the gravity measure, one has to quotient by these diffeomorphisms. If the background metric and dilaton profile have isometries, then some of the CKVs will be Killing vectors -- to quotient by these one divides the path integral explicitly by a factor of $\text{vol}(G)$ where $G$ is the isometry group. But the CKVs that are {\it not} Killing vectors show up as zero modes in the spectrum of the quadratic operator $\mathcal{D}$. One can quotient by the corresponding diffeomorphisms by omitting these zero modes from the path integral.

For the dilaton gravity examples we study, there is always a Killing vector corresponding to the $\textrm{U}(1)$ symmetry in the Euclidean time direction. The lukewarm solutions do not have further isometries and so we expect
\begin{equation}\label{lukewarmzero}
    \textrm{lukewarm}: \quad\quad\quad  n_{\rm zero}  = 5\,.
\end{equation}
On the other hand, the charged Nariai solution has a constant dilaton profile and a round sphere metric, so it has three Killing vectors, which leaves us with three remaining CKVs
\begin{equation}\label{nariaizero}
    \textrm{charged Nariai}: \quad  n_{\rm zero}  = 3\,.
\end{equation}
Both (\ref{lukewarmzero}) and (\ref{nariaizero}) are correct away from their intersection at the special point $Q= \frac{1}{4}$. At this point, we will see that there is an extra physical zero mode, so
\begin{equation}\label{lukewarmzero2}
   Q = \frac{1}{4}: \quad\quad\quad\quad \,\, n_{\rm zero}  = 6\,.
\end{equation}

\subsection{Numerical analysis using a discrete mesh}\label{sec:mesh}
In this section, we report the numerical results of the mesh method for the phase of the partition function of lukewarm and Nariai black holes. We will first obtain the spectrum of the quadratic operator $\mathcal{D}$ in~\eqref{quad}\footnote{\label{footnote5}We compute eigenvalues with respect to the naive inner product $\frac{1}{4G}\int \d^2x\sqrt{\hat g}(\varphi_1\varphi_2+\omega_1\omega_2)$. This differs from the inner product one gets by dimensionally reducing the higher dimensional inner product, but we spot-checked that this does not change the phase.} and then count the number of ``extra" negative modes after already accounting for one per mesh vertex
\begin{equation}\label{nminusdef}
n_{-} \equiv \text{(number of negative eigenvalues)}-\text{(total number of mesh vertices)}\,.
\end{equation}
The phase of the path integral will then be defined as $(-\i)^{n_-}$.

The mesh method, or finite element method (FEM)~\cite{ErnGuermond2004}, is a natural generalization of the finite difference formula $\partial_x^2F(x)\approx\frac{F(x+\delta x)+F(x-\delta x)-2F(x)}{(\delta x)^2}$ on a general geometry. The starting point is to generate a mesh on the sphere using e.g.~the packages \cite{gmsh,PerssonStrang2004}. For instance, we show a quasi-uniform mesh with $n_\text{vertices} = 695$ vertices (left) and a non-uniform mesh with $n_\text{vertices} = 598$ vertices (right):
\begin{equation}
     \includegraphics[width=0.5\textwidth, valign = c]{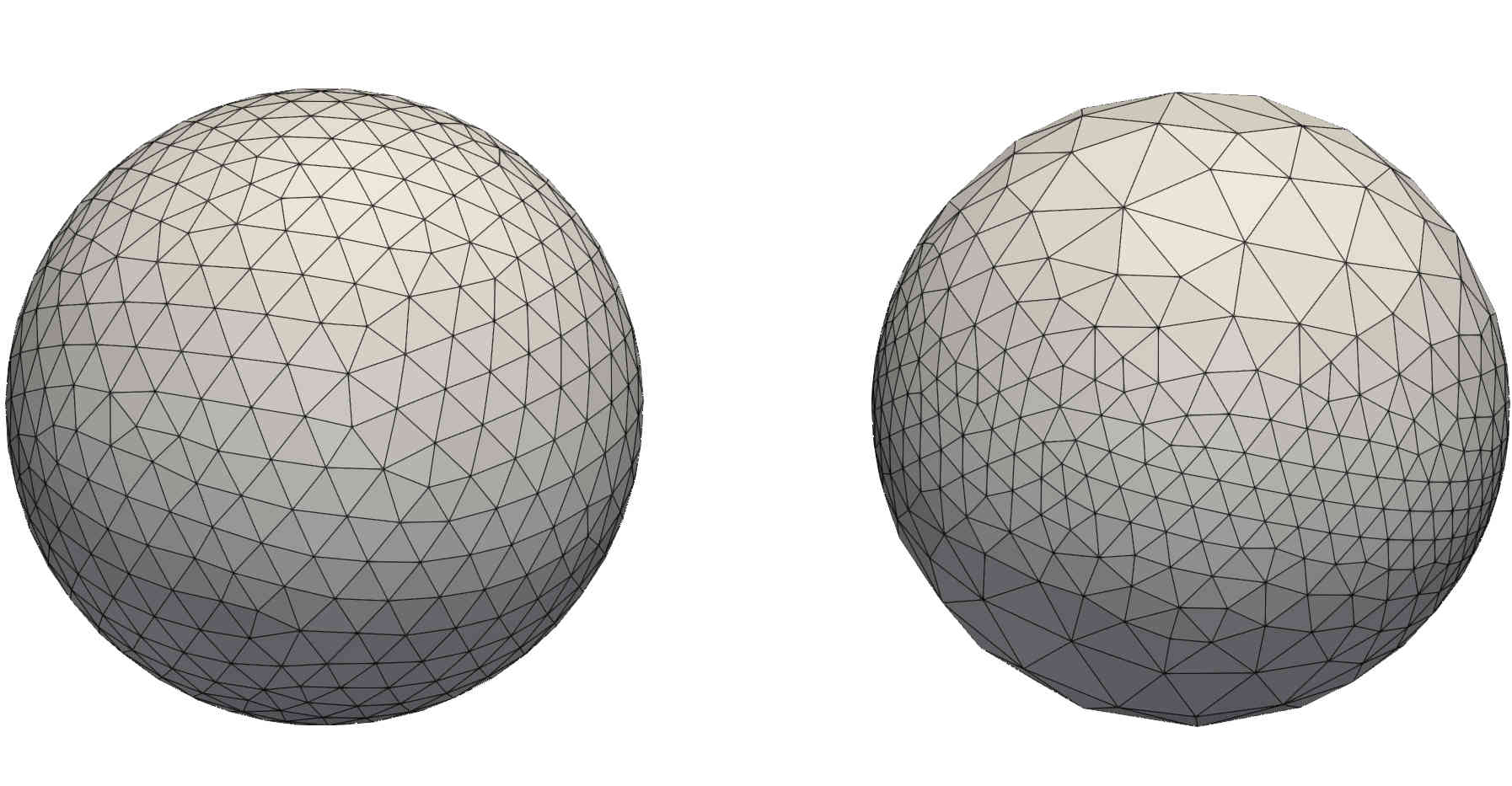}
\end{equation}
To use this mesh, it is convenient to write the solution (\ref{eq:  background metric}) as conformal to the round sphere, see (\ref{eq: coordinate trans2}), and use the mesh as an approximation to the round sphere. Software such as \cite{DBLP:journals/corr/abs-2109-12818} can then approximate the differential operator $\mathcal{D}$, as reviewed in \cref{app: mesh method detail}. Finally, the numerical spectrum of this operator is obtained as a generalized eigenvalue problem.

\paragraph{Lukewarm solutions:} From the numerically obtained spectrum, we indeed observe 5 approximate zero modes. After removing these and then computing $n_-$, we find $n_-=-3$ for sufficiently large $n_\text{vertices}$. This is true for all values of $Q$ that we checked. Note that for small values of $n_\text{vertices}$, we find different values for $n_-$ depending on details of the implementation. But for sufficiently large values we always find $n_- = -3$. In order to demonstrate the robustness of this conclusion, we perform the following three cross-checks.
\begin{itemize}
\item We tried different kind of triangulation (including non-uniform meshes) and different reference metrics, and we always found $n_-=-3$ for sufficiently large $n_\text{vertices}$.
\item We tried a completely different position space UV regularization scheme, the fuzzy sphere regularization. The idea is to use the small non-commuting-ness scale as a `covariant mesh'. There we also find the (regularized) number of negative modes are $n_-=-3$. We elaborate this method in \cref{app: fuzzy sphere method detail}. 

\item We checked that higher eigenvalues also appear to be converging (although slowly) to the exact answers. Here, the referee for the exact answers is the Chebyshev-Gauss-Lobatto collocation method, applied in each $U(1)$ symmetry sector. This method is extremely efficient for computing the spectrum, but it does not allow us to define $n_-$. We report the implementation of it in \cref{app: Chebyshev method detail}.
\end{itemize}

\paragraph{Charged Nariai solutions:}

For the thermodynamically unstable ($0<Q<\frac{1}{4}$) portion of the Nariai branch, we find $n_\text{zero}=3, n_-=0$. This matches onto results in \cite{Shi:2025amq,Ivo:2025yek} for the one-loop determinant for the uncharged Nariai black hole. On the other hand, for the stable ($\frac{1}{4}<Q<\frac{1}{2\sqrt{3}})$ portion, we find $n_\text{zero}=3, n_-=-3$. This matches onto an analytical computation in the ultracold limit below, see \cref{sec:ultracold}.

\paragraph{What happens at $Q = 1/4$:} The Nariai solutions have constant dilaton $\phi = \phi_0$ with $U(\phi_0) = 0$ and $U'(\phi_0) < 0$, and the metric is a round sphere with radius $\sqrt{-2/U'(\phi_0)}$, see \cref{fig:potential}. Any solution of this type will have at least three zero modes, corresponding to an $\ell = 1$ multiplet of CKV zero modes in the $\omega$ sector. However, if $U''(\phi_0)$ also happens to vanish, then there is also an $\ell = 1$ multiplet of physical zero modes in the $\varphi$ sector \cite{Maldacena:2019cbz,Mahajan:2021nsd,Ivo:2025yek}. For $Q < 1/4$, these become positive modes, and for $Q > 1/4$ they are negative. This explains the change in $n_-$, see \cref{app: nariai spectrum}. 

One can also approach $Q = 1/4$ from the lukewarm branch. The symmetry is enhanced at this point, so the spectrum organizes into $\text{SO}(3)$ multiplets. The $\ell = 1$ multiplet of physical $\varphi$ zero modes is assembled from two $m=\pm1$ would-be CKV zero modes , together with an $m = 0$\footnote{Here, $m$ is the $\text{U}(1)$ symmetry quantum number of lukewarm background.} mode  with eigenvalue that approaches zero, see \cref{app: Chebyshev method detail}. 

\paragraph{Summary: }We summarize the number of zero modes/negative modes, and phase of partition function of various cases:
\begin{equation}\label{table:nresult}
    \centering
    \begin{tabular}{|c|c|c|c|}
     \hline
     type & $n_\text{zero}$ & $n_-$ & phase of $Z$\\
     \hline
     lukewarm & $5$ & $-3$ &  $(-\i)^{-3} = -\i$\\
     \hline
     Nariai $Q<1/4$ & $3$ & $0$ &  $(-\i)^{0} = 1$\\
     \hline
     $Q = 1/4$ & $6$ & $-3$ &  ?\\
     \hline
     Nariai $Q > 1/4$ & $3$ & $-3$ &  $(-\i)^{-3} = -\i$\\
     \hline
    \end{tabular}
\end{equation}
\noindent Note that except for the case $Q = 1/4$, all zero modes are pure gauge, arising from conformal Killing vectors. For the case $Q = 1/4$ we have a physical zero mode, so the one-loop determinant diverges and the phase requires further study.

\subsection{A model for the lukewarm black holes}\label{sec:3dint}
In this section, we discuss a reduction of the lukewarm dilaton gravity theory to a three-dimensional integral that gives an interpretation for the phase. As motivation for this, one could optimistically model de Sitter plus a black hole by a pair of thermodynamic systems coupled by an integral over their shared inverse temperature $\beta$:
\begin{equation}\label{model1}
    Z = \int \d E_b \d\beta \d E_c e^{S_b(E_b) + S_c(E_c) - \beta(E_b + E_c)}, \hspace{20pt} E_b = \text{``BH energy''}, \hspace{20pt} E_c = \text{``cosmic energy''}.
\end{equation}
To derive something like this within the dilaton gravity system, we can look for a three-parameter off-shell configuration. Specifically, we take a $\text{U}(1)$ symmetric configuration where the dilaton is fixed to values $\phi_b,\phi_c$ at the two horizons, and the length of the curve at some intermediate value $\phi_g$ is also fixed to value $\ell$:
\begin{equation}
    \includegraphics[valign = c]{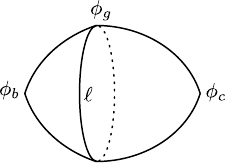}
\end{equation}
Then the model is
\begin{equation}\label{model2}
    Z = \int \d\phi_b \d\ell\d\phi_c e^{-I(\phi_b,\ell,\phi_c|\phi_g)}.
\end{equation}
Here $\phi_g$ stands for ``gluing'' since it is the value of $\phi$ where the two disks are glued together to make a sphere. The value of $\phi_g$ will not affect the qualitative features. In principle, this model ought to be supplemented by factors representing the one-loop integrals over all other modes, but we will see below that these factors will not contribute any net phase.

Let's work out the action $I(\phi_b,\ell,\phi_c|\phi_g)$ and study the one-loop determinant around a classical solution. The action is the sum of two contributions, from the two disks that are glued together. For each disk, the action is
\begin{equation}
    I = -\frac{1}{4G}\int \d^2 x \sqrt{g}(\phi R + U(\phi)) - \frac{1}{2G}\int_{\text{bdy}} \d y \sqrt{h} \phi K.
\end{equation}
The solution is (\ref{eq:  background metric}) where for the (left) disk surrounding the black hole horizon, 
\begin{align}
    A(\phi) &= \int_{\phi_b}^\phi \d\phi U(\phi)\\
\sqrt{g}R &= -A''(\phi) + \delta^2(x - \text{horizon})\left(4\pi - \beta U(\phi_b)\right), \hspace{20pt} K = A'(\phi) / 2\sqrt{A(\phi)}.
\end{align}
Note that the off-shell ansatz allows delta function curvature at the two horizons. Plugging in, the action is
\begin{align}
    2G\cdot I(\phi_b,\ell|\phi_g) &=  - \beta \phi_g\frac{U(\phi_g)}{2} - 2\pi\phi_b+ \frac{\beta}{2}\phi_b U(\phi_b)-\frac{\beta}{2}\int_{\phi_b}^{\phi_g}\d\phi\left(-\phi A''(\phi) +A'(\phi)\right)\\
    &=-2\pi\phi_b -\ell \sqrt{\int_{\phi_b}^{\phi_g}\d\phi U(\phi)}.
\end{align}
After recognizing the integrand as a total derivative, $\partial_\phi(-\phi A' + 2A)$, most terms canceled, and in the final line $\ell = \sqrt{A(\phi_g)}\beta$ was used. After working out a similar contribution for the (right) disk surrounding the cosmic horizon and adding the two together, one finds
\begin{align}
    2G \cdot I(\phi_b,\ell,\phi_c|\phi_g) &= -2\pi(\phi_b+\phi_c) - \ell\left(\sqrt{\int_{\phi_b}^{\phi_g}\d\phi U(\phi)} - \sqrt{-\int_{\phi_g}^{\phi_c}\d\phi U(\phi)}\right).
\end{align}
The equations of motion for $\phi_b,\phi_c$ imply that the two horizons should be smooth, and the equation of motion for $\ell$ implies $\int_{\phi_b}^{\phi_c}\d\phi U(\phi) = 0$. 

To study the one-loop determinant around such a solution, we need to compute the nonzero second derivatives of the action, which are
\begin{align}
    \partial_{\phi_b}^2 I = \pi\frac{U'(\phi_b)+\frac{8\pi^2}{\ell^2}}{GU(\phi_b)}, \hspace{20pt} 
    \partial_{\phi_c}^2 I = \pi\frac{U'(\phi_c)+\frac{8\pi^2}{\ell^2}}{GU(\phi_c)}, \hspace{20pt}
    \partial_{\phi_b}\partial_\ell I &=\partial_{\phi_c}\partial_\ell I = \frac{\pi}{G\ell}.
\label{eq: 4.29}
\end{align}
The explicit form of the one-loop determinant is
\begin{align}\label{woeiru}
    Z &= e^{S}\int_{-\infty}^\infty \d(\delta\phi_b)\d(\delta\phi_c)\d(\delta \ell) \exp\left[-\frac{1}{G}\left(c_b (\delta \phi_b)^2 + c_c(\delta \phi_c)^2 + c\,\delta\ell(\delta\phi_b + \delta \phi_c)\right)\right].
\end{align}
For the lukewarm black holes under consideration, $c_b >0$, $c_c < 0$, and $c_b + c_c > 0$. The fluctuation matrix has one negative eigenvalue, predicting the phase $-\i$ for the path integral.\footnote{Although it doesn't arise for the specific case of the charged black holes in dS$_4$, one can construct dilaton potentials for which the lukewarm solutions would be thermodynamically unstable, see \cref{app: unstable lukewarm}. In that case, $c_b + c_c < 0$ and there would be two negative eigenvalues and phase $(-\i)^2$. This agrees with the mesh numerics for such cases. One reason why this simple integral correctly captures the phase of both stable and unstable lukewarm situations is that the mode crossing zero, responsible for $n_-$ jumping from $-3$ to $-2$, lives in the $m=0$ sector, so a mini-superspace model is enough. To cook up a toy model explaining the phase jump  of the Nariai branch, this would not be enough since those three modes form an $\ell = 1$ multiplet with $m=-1,0,1$, see \cref{app: nariai spectrum}.} Crucially, this matches the phase of the full one-loop determinant. This gives an interpretation of the phase as arising from the instability of the $\delta\phi_c$ integral -- the fluctuation in the area of the cosmic horizon. 

It is interesting to compare to the computation of the Euclidean Schwarzschild black hole in 4D flat space. This has a negative mode in the canonical ensemble \cite{Gross:1982cv,PhysRevD.30.1153,PhysRevD.18.2733} that can be explained by a similar integral, over the area-radius of the horizon:\footnote{See appendix F.1 of \cite{Maldacena:2019cbz} and more recently \cite{Mahajan:2025bzo,Ailiga:2025osa} for a similar discussion of small black holes in AdS.}
\begin{equation}\label{schcano}
    Z_{\text{4D Sch}}(\beta) = \int \d r_h \exp\left[-\frac{1}{G}\left(\beta \frac{r_h}{2} -\pi r_h^2\right)\right].
\end{equation}
As in (\ref{woeiru}), the negative mode is contained in the integral itself -- the integrand is positive. A difference is that for the flat-space Schwarzschild black hole the integrand is positive because there are no other negative modes after Wick-rotation of the conformal factor. In the de Sitter case, the integrand is positive because there are {\it minus four} other negative modes and $(-\i)^{-4} = 1$.

\subsection{Ultracold Nariai}\label{sec:ultracold}
In this section, we study analytically the phase in the ultracold Nariai limit:
\be
Q\rightarrow {1\over 2\sqrt{3}},~~~\phi\rightarrow {1\over 6}.
\ee
This is a limit where classically the de Sitter horizon is in thermal equilibrium with an extremal black hole, and so the system approaches zero temperature and minimum entropy. 
Interestingly, in this limit, the black hole horizon and the de Sitter horizon become strongly coupled to each other. More precisely, as we approach the ultracold point, one of the $S^2$ factors in the Nariai $S^2\times S^2$ geometry becomes large and develops strong quantum fluctuations in its shape, somewhat like the enhancement of the Schwarzian fluctuations of near-extremal black holes.

This quantum fluctuation can be described by the dilaton gravity theory. In the ultracold limit, the dilaton field $\phi$ is almost a constant, so one can consider the fluctuation of its deviations $\varphi$. The dilaton potential expanding around this point has vanishing derivative, and so one has to keep the next quadratic term $\propto \varphi^2$. So in the end, the soft modes are controlled by a $\varphi^2$ dilaton theory that we will now analyze.

More explicitly, let's consider the following small $q$ expansion of the dilaton action \eqref{eqn:dilatonac} up to quadratic order in $q$:
\bea
Q&=&\sqrt{\frac{1}{12}-3 q^2},~~~\phi={1\over 6}+ q\varphi,~~~g_{\mu\nu}={\tilde g_{\mu\nu}\over c q},~~~U(\phi)\approx c q^2(1-\varphi^2);~~~\\
I&\approx&-S_0\chi(M)-{q\over 4G} \int \sqrt{\tilde g}(\varphi \tilde R+1-\varphi^2),~~~S_0={\pi\over 6 G};~~~c=36\sqrt{6}.
\label{eq: ultracold scaling}
\eea
Here $S_0$ is the sum of the classical entropies of the two horizons. 
We have scaled both the dilaton variation and the 2D metric by a power of $q$ such that the classical saddle of the rescaled field variables takes order one value. In particular, this means a local temperature $T_{\text{loc}}$ measured by some static bulk observers living in the classical geometry scales as $q^{1/2}$, see~\eqref{eq: ultracold temperature}. This in turn tells us that the coupling constant scales as $q^{-1}\sim T_{\text{loc}}^{-2}$. The theory becomes strongly coupled as we approach the ultracold limit, leading to logarithmic corrections to the entropies of the de Sitter and black hole horizons. The $\varphi$ integral is Gaussian. After integrating it out, we obtain
\be\label{eqn:R2ac}
I=-S_0 \chi(M)-{q\over 16 G} \int \sqrt{\tilde g}(\tilde R^2+4).
\ee
From this $R^2$ action, it is clear that the path integral of this theory is not convergent when we sum over real Euclidean geometries, which is a type of IR instability of sphere path integral. In fact, instead of being bounded from below, the action is bounded from above.\footnote{In the strict $q=0$ limit, the action is just a constant. Perhaps one can understand this in terms of the black hole stabilizing the de Sitter horizon. It would be interesting to understand this ``topological'' limit.}

Let's now analyze the phase of the $S^2$ saddle and the logarithmic correction of this theory. Expanding the action \eqref{eqn:R2ac} around the unit round sphere in the conformal gauge, $\tilde{g}=e^{2\omega } \tilde g_{S^2}$, one finds the quadratic action is proportional to
\be
I\propto-{q^2\over G} \sum_{\ell,m}\left(1-{\ell(\ell+1)\over 2}\right)^2\omega_{\ell,m}^2.
\label{eq: ultracold spectrum}
\ee
Here, $\omega_{\ell,m}$ are the spherical harmonic modes normalized using the physical metric $g$, not the metric $\tilde{g}$. They are all unstable except the three $\ell=1$ zero modes, corresponding to the conformal Killing modes on the sphere (see \eqref{nariaizero}).\footnote{The one-loop action in timelike Liouville is similar, but there the $\ell =0$ mode is positive \cite{Anninos:2021ene}.} The phase of the partition function comes from rotating all the unstable modes, each one gives a factor of $-\i$. Due to the three missing zero modes, that leads to $\i^3$. The three zero modes also lead to a $q^{3}$ one-loop determinant. All together,
\be
Z\propto \i^3 q^{3} e^{S_0}\propto \i^3 T_{\text{loc}}^{6} e^{S_0}.
\ee
The $\i^3 = -\i$ phase agrees with our numerical results for the $Q > 1/4$ Nariai branch. The $T_{\text{loc}}^{6}$ indicates a decrease in the gravitational entropy due to quantum fluctuations of the soft modes. It would be interesting to understand whether in a full computation (as opposed to one-loop) the entropy vanishes in the limit of $q\rightarrow 0$, as in the case of non-supersymmetric JT.

\section{The phase of the state-counting partition function}\label{sec:statecount}
In the previous sections, we studied the phase of the Euclidean gravity path integral. In this section, we compute the phase of the state-counting partition function. The difference is that the Euclidean gravity path integral is over real Euclidean geometries (up to contour rotations for convergence), whereas for the state-counting partition function we modify the integral in order to impose the Hamiltonian constraint $H = 0$ \cite{Casali:2021ewu, Marolf:2022ybi, Held:2024rmg,DeVuyst:2024fxc,Held:2025mai,Maldacena:2024spf}.
\begin{align}
    \text{naive Euclidean gravity path integral $Z$:}& \hspace{20pt} \int_0^\infty \d\beta\\
    \text{state-counting partition function $Z_{\text{Count}}$:}& \hspace{20pt} \int_{-\i\infty}^{\i\infty} \frac{\d\beta}{2\pi\i}.\label{ilt}
\end{align}
This contour is familiar from the inverse Laplace transform that computes the density of states $\rho(E)$ for e.g.~AdS black holes, and it is natural from the Lorentzian point of view \cite{Marolf:2022ybi,Dittrich:2024awu,Held:2025mai}.

\newcommand{\Eo}{{\boldsymbol E}}
\newcommand{\rhoo}{{\boldsymbol \rho}}
\subsection{dS with an observer}\label{sec:dS3phase}
Let's temporarily put charged black holes aside and consider the case of dS with a probe observer, starting with the case of dS$_3$. We will imagine that all but three modes of the path integral have already been integrated out in a one-loop approximation, leaving the energy of the observer and the two defect angles around the observer's worldline and the horizon. The off-shell metric is
\be
\d s^2 = \cos^2(\theta) \d\tau^2+\d\theta^2 + \sin^2(\theta) \d \alpha^2, \hspace{20pt} \tau\sim\tau+\beta, \hspace{20pt} \alpha\sim \alpha+A,
\ee
where the observer sits at $\theta = 0$. The Einstein-Hilbert action is
\begin{align}
-I = \frac{1}{16\pi G}\int \sqrt{g}(R-2\Lambda) &=-\frac{\beta A}{8\pi G}+ \frac{\beta+A}{4G}.
\end{align}
Coupling to a probe observer with energy $\Eo$ and density of states $\rhoo$, the model for the Euclidean gravity path integral is
\begin{equation}
    Z = \int \d \Eo \d\beta \d A \ e^{-\beta \Eo - I}\rhoo(\Eo).
\end{equation}
Expanding $A = 2\pi + \delta A$ and $\beta = 2\pi + \delta\beta$, we get
\begin{equation}\label{Z3d}
    Z = e^{\frac{\pi}{2G}}\int \d \Eo \d(\delta\beta) \d (\delta A) e^{- 2\pi \Eo -\delta\beta(\Eo + \frac{\delta A}{8\pi G})}\rhoo(\Eo).
\end{equation}
For fixed $\Eo$, there is one negative mode in the $\delta\beta,\delta A$ sector, reproducing the phase $(-\i)$ of the Euclidean gravity path integral with the observer included \cite{Maldacena:2024spf}. This means that the modes we have retained explicitly already produce the correct phase, and therefore the contribution of all of the other modes will be to multiply this integral by a positive and decoupled one-loop determinant. 

After modifying the $\beta$ integral for the state-counting partition function, the answer is real and positive
\begin{align}
    Z_{\text{Count}} &=  \int\d \Eo \d A \int_{-\i\infty}^{\i\infty} \frac{\d\beta}{2\pi\i}e^{-\beta \Eo - I}\rhoo(\Eo)\\
    &=e^{\frac{\pi}{2G}}\int \d \Eo e^{- 2\pi \Eo}\rhoo(\Eo).
\end{align}

In higher dimensions, the analysis is similar, although differing in two respects. First, in order to avoid forming a black hole, the observer needs a nonzero size. In \cref{app:highD} we study $D = 4$ and model the the observer as a perfect fluid occupying the center of the static patch. Second, the structure of the integral differs from (\ref{Z3d}) in that there is a further term in the action analogous to $(\delta A)^2$. This generates a quadratic term in $\delta\beta$ after doing the integral. If $\beta$ is defined as the length of the observer's worldine, then the $(\delta A)^2$ term is a ``wrong-sign'' Gaussian, leading to a ``right-sign'' Gaussian for $\delta\beta$ \cite{Maldacena:2024spf}. From our perspective this doesn't make much difference, and the final $\Eo$ integrand remains positive.

\subsection{The lukewarm black hole}\label{subsec:lukewarm}
Let's now study this procedure in the dilaton gravity reduction of the lukewarm black hole, using the three-dimensional integral representation (\ref{model2}). The one-loop Euclidean gravitational path integral is (after omitting some positive constants)
\begin{equation}\label{intmodel4.2}
    Z = e^{S}\int_{-\infty}^\infty \d(\delta\phi_b)\d(\delta\phi_c) \d(\delta \ell) e^{-c_b (\delta \phi_b)^2 - c_c(\delta \phi_c)^2 - \delta\ell(\delta\phi_b + \delta \phi_c)}.
\end{equation}
The variables $\delta \phi_b, \delta\phi_c$ are small fluctuations in the value of the dilaton at the black hole horizon and cosmic horizon, and $\ell$ is the length of an intermediate curve at some particular fixed value of the dilaton. The coefficient $c_b$ is positive and $c_c$ is negative, with $c_b + c_c >0$.

The integral (\ref{intmodel4.2}) has one negative mode, reproducing the phase $(-\i)$ of $Z$. For the state-counting partition function, we interpret $\ell$ as $\beta$ and modify the integral to
\begin{align}\label{woeiru}
    Z_{\text{Count}} &= e^{S}\int_{-\infty}^\infty \d(\delta\phi_b)\d(\delta\phi_c) \int_{-\i\infty}^{\i\infty} \frac{\d(\delta \ell)}{2\pi\i} e^{-c_b (\delta \phi_b)^2 - c_c(\delta \phi_c)^2 - \delta\ell(\delta\phi_b + \delta \phi_c)}\\
    &=e^{S}\int_{-\infty}^\infty \d(\delta\phi_b)e^{-(c_b+c_c)(\delta \phi_b)^2}.
\end{align}
The final integrand is positive. The integral itself is convergent for the stable case $c_b + c_c > 0$.\footnote{For the unstable case (see  \cref{app: unstable lukewarm}) the integral would diverge in the one-loop approximation, and in this case it might be appropriate to define the integral with a cutoff rather than a contour rotation.}

\subsection{On the overall sign}
Our results differ by a sign relative to \cite{Maldacena:2024spf}. Let's review the computation of $\mathcal{Z}_\text{Count}$ in section 4.2 of that work. The starting point was a representation of $\mathcal{Z}_{\text{Obs}}=Z$ in which all modes were integrated out except $\delta\beta$. We omit positive constants and refer to $\delta\beta$ as $\beta$ to simplify the equations:
\begin{equation}\label{Z6}
    \mathcal{Z}_\text{Obs} \sim (-\i)\int_{-\infty}^{\infty}  \d\beta e^{-\beta^2} \sim (-\i).
\end{equation}
The explicit phase $(-\i)$ comes from the integral over other modes. The stable Gaussian for fluctuations in $\beta$ is a bit unfamiliar, corresponding to a thermodynamic system with negative specific heat. In \cite{Maldacena:2024spf} this was deduced from the fact that $\langle (\delta\beta)^2\rangle > 0$. Within the models we studied, one can see it by integrating out $\phi_b,\phi_c$ from (\ref{intmodel4.2}) with $\ell\to\beta$ or integrating out $\mu$ from (\ref{intoutmu}). More generally, the sign is related to the fact that if matter is added to the static patch, the length of the observer's worldline gets shorter (for $D > 3$). 

To compute $\mathcal{Z}_\text{Count}$, one should replace the $\beta$ integral by an inverse Laplace transform contour. Along {\it that} contour, the $\beta$ integral is divergent. The prescription in \cite{Maldacena:2024spf} is to rotate back to the real axis clockwise, avoiding the line of maximal growth:
\begin{align}
\mathcal{Z}_\text{Count} &\sim (-\i)\int_{-\i\infty}^{\i\infty}  \frac{\d\beta}{\i} e^{-(1-\i\epsilon)\beta^2} \\ &\to (-\i)^2\int_{-\infty}^\infty \d\beta e^{-\beta^2}\\ &\sim -1.
\end{align}
Roughly, the variable $s = -\i\beta$ is unstable, so it contributes a factor of $(-\i)$. Combining with the original $(-\i)$ from the other modes, one finds $(-\i)^2 = -1$. 

We suggest a different prescription, based on the idea that the $\beta$ integral is supposed to impose the Hamiltonian constraint. To implement this, it is necessary to expose at least one further integral, over the energy variable conjugate to $\beta$. Let's call this variable $E$. The integral over $\beta,E$ is conditionally convergent if the $\beta$ integral is done first, imposing a constraint $\delta(E)$.\footnote{The importance of doing the inverse Laplace transform integral over $\beta$ first was emphasized in \cite{Marolf:2022ybi}.} The final answer is positive. To illustrate this in the simplest context, let's arrange an action for $\beta$ and $E$ so that (\ref{Z6}) arises from integrating out $E$: 
\begin{equation}
    Z \sim \int_{-\infty}^\infty \d\beta\d E e^{(1-\i\epsilon)\left(-\beta E + E^2\right)}\sim (-\i)\int_{-\infty}^\infty \d\beta e^{-(1-\i\epsilon)\beta^2}.
\end{equation}
Our prescription for $Z_\text{Count}$ would then be\footnote{Let's explain the relationship between this two-dimensional integral and our previous models. In each case the idea is that $E$ is whatever $\delta\beta$ multiplies in the action. So, in the dS$_3$ example from \cref{sec:dS3phase}, $E= \Eo +\delta A/8\pi G$ (this example is a bit special because there is no quadratic term for $E$). In the dS$_4$ example from \cref{app:highD}, $E = \Eo - \mu$. In the dilaton gravity example from \cref{subsec:lukewarm}, $E = \delta\phi_b  +\delta\phi_c$.}
\begin{align}
    Z_\text{Count} &\sim \int_{-\infty}^\infty\d E\int_{-\i\infty}^{\i\infty} \frac{\d\beta}{2\pi\i} e^{-\beta E + E^2}\\
    &= \int_{-\infty}^\infty\d E\delta(E)e^{E^2}\\
    &= 1.
\end{align}
See \cref{app:2dintegral} for more on this integral. As a sanity check, note that this particular integral arises in computing the microcanonical density of states for the flat-space Schwarzschild black hole in 4D, i.e.~the inverse Laplace transform of (\ref{schcano}).

\section{Discussion}

In this paper we followed a suggestion in \cite{Maldacena:2024spf} to use charged black holes as observers in de Sitter space, in order to understand the phase of the sphere partition function. Within a dilaton gravity reduction, we found the same phase $\i^3 = -\i$ as in \cite{Maldacena:2024spf} for stable de Sitter black holes. For the unstable Nariai solution, we found trivial phase consistent with the results of \cite{Ivo:2025yek,Shi:2025amq}. Our main technique to compute the phase was numerical, but we also studied the ultracold Nariai limit analytically. In this limit the black hole horizon and the de Sitter horizon become strongly fluctuating, governed by a $R^2$ two-dimensional gravity theory. We proposed a slightly different prescription from \cite{Maldacena:2024spf} for relating the Euclidean gravity path integral to a state-counting partition function. With this change, the density of states is real and positive, although this relies on the positivity of an integral with minus four negative modes: $(-\i)^{-4} = 1$. It would be nice to have a more complete understanding of this factor.

Together with the von Neumann algebra discussion in \cite{Chandrasekaran:2022cip} and the double cone wormhole \cite{Yang:2025lme},\footnote{See also \cite{Penington:2019kki,Chen:2020tes,Mirbabayi:2023vgl,Fumagalli:2024msi} for preliminary studies of de Sitter wormhole effects and \cite{Usatyuk:2024mzs,Harlow:2025pvj,Abdalla:2025gzn} for related puzzles.} a positive state-counting partition function provides nontrivial evidence that the de Sitter horizon, as seen by an observer, admits a quantum mechanical description.

Clearly, many more checks must be performed to substantiate this. For example, one could study $G_N$ corrections to the existing theoretical evidence. It also seems important to understand the negative OTOC/signalling puzzle \cite{Gao:2000ga,Anninos:2018svg,Aalsma:2020aib,sekinoSusskind}, the puzzle of the one-sided observer discussed in \cite{Chandrasekaran:2022cip}, and more generally, to analyze the potential instabilities of the sphere path integral in the presence of observers, aiming to develop a mathematically consistent prescription at the nonperturbative level.
Ref.~\cite{Witten:2023xze} already proposed a framework that, in principle, allows one to incorporate both perturbative and nonperturbative corrections. The proposal is that, when we focus on states defined in the presence of an observer, the Hartle–Hawking state becomes a natural state of maximal entropy in quantum gravity. Given the many lessons learned by studying near-extremal black holes in other settings, it might also be interesting to further study the ultracold limit, which is a natural state of minimal entropy.

\section*{Acknowledgements}
We are grateful to Frank Ferrari,  Xuyao Hu, Victor Ivo, Albert Law, Guanda Lin, Juan Maldacena, Don Marolf, Henry Maxfield, Xiao-Liang Qi, Steve Shenker, and Zimo Sun for discussions. This work was supported in part by DOE grants DE-SC0021085 and DE-SC0026143 and by a grant from the Simons foundation (926198, DS).
ZY is supported by NSFC Grant No. 12447108, 12342501, 12475071. HT is supported by Shoucheng Zhang Fellowship. Part of this work was performed during the Aspen summer program ``Recent developments in String Theory,'' supported by National Science Foundation grant PHY-2210452.

\appendix

\section{Negative modes and \texorpdfstring{$\pm\i$}{pm i}}\label{app:negative}
Consider the integral
\begin{equation}
    f(a) = \int_{-\infty}^\infty \frac{\d x}{\sqrt{\pi}}e^{-a x^2} = \frac{1}{\sqrt{a}}, \hspace{20pt} \text{Re}(a)>0.
\end{equation}
This expression can be continued to negative $a$, either through the upper half-plane or the lower half-plane, giving answers that differ by a sign. It is important to pick a consistent choice, and we will follow the convention in \cite{Maldacena:2024spf} where the action is multiplied by $(1-\i \epsilon)$ and the wrong-sign Gaussian integral is then defined as
\begin{align}
    \int \frac{\d x}{\sqrt{\pi}}e^{x^2} \to \int \frac{\d x}{\sqrt{\pi}}e^{(1-\i\epsilon)x^2} = f(\i\epsilon - 1) \equiv (-\i),
\end{align}
where $\epsilon > 0$ and $f$ is continued the ``short way'' through the upper half-plane.

\section{A two dimensional integral}\label{app:2dintegral}
Consider the integral
\begin{equation}
    \int_{-\infty}^\infty \frac{\d s \d E}{2\pi}e^{\i s E + E^2}.
\end{equation}
In the main text, we argued that the ``right'' definition is as a conditionally convergent integral with no contour rotations:
\begin{align}
    \int_{-\infty}^\infty \d E\int_{-\infty}^\infty \frac{\d s}{2\pi}e^{\i s E + E^2} = \int_{-\infty}^\infty \d E\delta(E)e^{E^2} = 1.
\end{align}
As an alternative, one could define the integral by analytic continuation, starting from a convergent integral. Depending on how this is done, one can get different answers, corresponding to different contour rotations. In particular, if we introduce a parameter $a$ and start near $a = -1$ then the following integrals converge
\begin{align}
    \int_{-\infty}^\infty \frac{\d s \d E}{2\pi}e^{a(\i s E + E^2)} &= \frac{1}{-a}\\
    \int_{-\infty}^\infty\frac{\d s \d E}{2\pi}e^{\i s E + aE^2} &= 1.
\end{align}
Continuing the first expression to $a = 1$ gives a negative answer as in \cite{Maldacena:2024spf}, and continuing the second expression (trivially) to $a = 1$ gives a positive answer.

\section{Details on the numerical methods}
\subsection{Method I: discrete mesh}
\label{app: mesh method detail}
In this appendix we elaborate on how to practically implement the mesh method to obtain the spectrum of one-loop fluctuations in conformal gauge, studied in \cref{sec:mesh}. We first recall the eigenvalue equation:
\begin{equation}\label{eigeqn}
\begin{pmatrix}  - U''/2   &    \hat{\nabla}^2  -  U' \\
     \hat{\nabla}^2    -  U' & -2 U  \end{pmatrix}
\begin{pmatrix}
\varphi\\
\omega
\end{pmatrix}=
\lambda\begin{pmatrix}
\varphi\\
\omega
\end{pmatrix}
\end{equation}
where $\hat\nabla^2$ is the Laplacian on the metric of the classical solution~\eqref{eq:  background metric}. It is convenient to choose a reference metric as the unit sphere, for which a quasi-uniform mesh can be readily generated. The coordinate transformation to go to the reference metric is given by 
\begin{equation}
\frac{\d\phi^2}{A(\phi)}+A(\phi)\d\tau^2=\Omega(\theta)^2\left( \d\theta^2+\sin^2\theta\d\alpha^2\right), \quad  \alpha\sim\alpha+2\pi,\ 
\end{equation}
where $\Omega^{-2}=(\partial_{\phi}\theta)^2A(\phi) $ with the inverse transform $\theta({\phi})$ given explicitly by
\begin{equation}
\begin{aligned}
\theta(\phi)&=\pi-\arccos\left\{\tanh\left[\operatorname{arctanh}\left(\frac{2\sqrt{\phi}-1}{\sqrt{1-4Q}}\right) + \sqrt{\frac{1-4Q}{1+4Q}}\operatorname{arccoth}\left( \frac{2\sqrt{\phi}+1}{\sqrt{1+4Q}} \right)\right]\right\}\,.\\
\end{aligned}
\label{eq: coordinate trans2}
\end{equation}
We can rewrite $\hat\nabla^2=\Omega^{-2}\nabla^2_{S^2}$ using the unit round sphere metric Laplacian $\nabla_{S^2}^2$. It is convenient to multiply $\Omega^2$ on both sides of (\ref{eigeqn}) and obtain:
\begin{equation}\label{disteqn}
\begin{pmatrix}  - \Omega^2U''/2   &    \nabla^2_{S^2}  -  \Omega^2U' \\
     \nabla^2_{S^2}    -  \Omega^2U' & -2 \Omega^2U  \end{pmatrix}
\begin{pmatrix}
\varphi\\
\omega
\end{pmatrix}=
\lambda\cdot\Omega^2\begin{pmatrix}
\varphi\\
\omega
\end{pmatrix}
\end{equation}
Now, we denote $i=1,...,n_{\text{vertices}}$ as index of mesh vertices points, and expand wavefunction $\varphi(x^\mu),\omega(x^\mu)$ (here $x^\mu=(\theta,\alpha)$ denote the usual coordinates on $S^2$) on a set of basis functions $h_i(x^\mu)$:
\begin{equation}
\varphi(x^\mu)\equiv\sum_i\varphi_ih_i(x^\mu), \ \ \omega(x^\mu)\equiv\sum_i\omega_ih_i(x^\mu)
\end{equation}
There could be many choices of the basis functions. We nevertheless pick the simplest one - called the P1-linear function~\cite{ErnGuermond2004}, where $h_i=1$ at vertex $i$, and decrease linearly to zero at neighboring points while remaining zero everywhere else. Pictorially, one can visualize the basis function $h_i$ as a ``pyramid"  peaked at vertex $i$:
\begin{equation}    \includegraphics[width=0.4\linewidth]{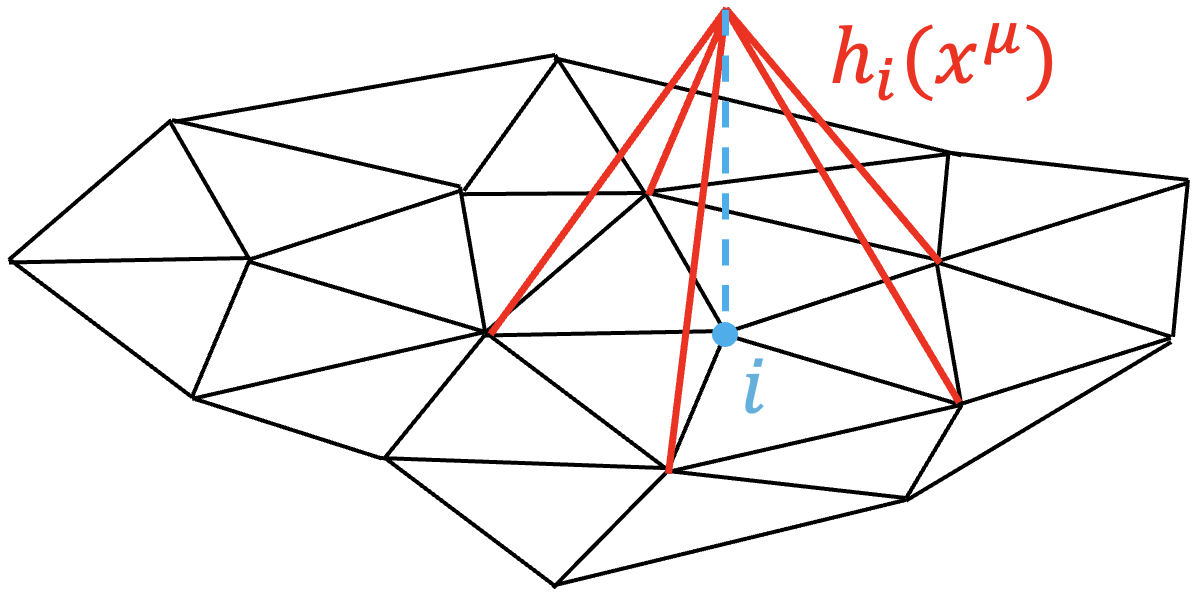}
\end{equation}
Then, we can calculate a $n_{\text{vertices}} \times n_{\text{vertices}}$  ``stiffness matrix" $K$ and a ``mass matrix" $M[f]$~\cite{ErnGuermond2004}, whose matrix elements are the Laplacian and scalar function $f(x^\mu)$ sandwiched between basis functions:
\begin{equation}
K_{ij}\equiv-\int_{S^2}\d^2x\sqrt{g_{S^2}}(\nabla_{S^2}h_i)\cdot(\nabla_{S^2}h_j),\ \ M[f]_{ij}\equiv\int_{S^2} \d^2x\sqrt{g_{S^2}}(fh_ih_j)\,. 
\end{equation}
With these definitions, our discretized eigenvalue equation takes the form
\begin{equation}
\mathcal D_{\text{mesh}}\Psi_\text{mesh}=\lambda\mathcal M_{\text{mesh}}\Psi_\text{mesh}, \quad  \quad \Psi_{\text{mesh}}\equiv\begin{pmatrix}
\varphi_1, & \cdots & \varphi_{n_{\text{vertices}}} , & \omega_1, & \cdots & \omega_{n_{\text{vertices}}} 
\end{pmatrix}^\intercal ,
\end{equation}
where
\begin{equation}
\mathcal D_\text{mesh}\equiv
\begin{pmatrix}
-M[\Omega^2U'']/2 &  K-M[\Omega^2U'] \\
 K-M[\Omega^2U'] & -2M[\Omega^2U]
\end{pmatrix},\quad  \mathcal M_\text{mesh}\equiv M[\Omega^2]
\end{equation}

We now report aspects of the spectrum of the lukewarm branch found in numerics. Considering $Q=0.2$ as an example, we show the spectral data for various mesh sizes in  \cref{table:meshdata}. In the table, ``zero mode error" records the maximal absolute value of the five nearly zero eigenvalues, $\lambda_1$ records the value of the first non-zero eigenvalue. 
\begin{table}[t]
  \centering
  \begin{minipage}[t]{0.48\linewidth}
    \centering
\begin{tabular}{|c|c|c|c|}
    \hline
     $n_{\text{vertices}}$ & $n_-$ & zero mode error & $ \lambda_1$  \\

     \hline
     $206$ & $-3$ & $0.8237$ & $3.9189$\\
     
     \hline
     $777$ & $-3$ & $0.2095$ & $3.8396$\\

     \hline
     $1540$ & $-3$ & $0.1043$ & $3.8238$\\

     \hline
     $3012$ & $-3$ & $0.0529$ & $3.8154$\\
     \hline
     $8222$ & $-3$ & $0.0192$ & $3.8099$\\

     \hline
     $18389$ & $-3$ & $0.0085$ & $3.8080$\\

     \hline  
    \end{tabular}
  \end{minipage}\hfill
  \begin{minipage}[t]{0.48\linewidth}
    \centering
\begin{tabular}{|c|c|c|c|}
    \hline
     $n_{\text{vertices}}$ & $n_-$ & zero mode error & $ \lambda_1$  \\

     \hline
     $211$ & $-3$ & $0.8752$ & $3.8897$\\
     
     \hline
     $777$ & $-3$ & $0.2018$ & $3.8345$\\

     \hline
     $1526$ & $-3$ & $0.0989$ & $3.8215$\\

     \hline
     $3032$ & $-3$ & $0.0510$ & $3.8145$\\
     \hline
     $8249$ & $-3$ & $0.0188$ & $3.8095$\\

     \hline
     $18176$ & $-3$ & $0.0084$ & $3.8079$\\

     \hline  
    \end{tabular}
  \end{minipage}

  \caption{The numerical results for the $Q = 0.2$ lukewarm solution via the mesh method. \textbf{Left:} Use quasi-uniform mesh. \textbf{Right:} Use non-uniform mesh.} 
  \label{table:meshdata}
\end{table}

\subsection{Method II: fuzzy sphere regularization}
\label{app: fuzzy sphere method detail}

Here we study another numerical method, the fuzzy sphere regularization, which in a similar vein as the mesh method can unambiguously determine the (regularized) number of the negative mode in the spectrum. 

The fuzzy sphere method is motivated from studying the Landau problem on sphere: a free particle subject to the magnetic field sourced by a monopole (with quantized charge $S$) at the center of a sphere. Just as solving Landau problem on the plane, the center-of-mass position of the particle satisfy non-commuting algebra, living in a noncommutative geometry. The degree of noncommutativity is controlled by the cyclotron length $\sim S^{-1}$, which realizes the idea of putting a UV cutoff in the position space. So, the rotation of a ``whole field" is well defined in this setting. In the large $S$ limit,  we expect the spectrum converges to continuous limit.

At a technical level, the lowest landau level (LLL) furnishes a spin-$S$ rep of $SO(3)$. Define $[J_i,J_j]=i\varepsilon_{ijk}J_k$ as the usual $(2S+1)$ dimensional spin matrices. Now, the continuous scalar fields $(\varphi,\omega)$ becomes two $(2S+1) \times (2S+1)$ matrices~\cite{Balachandran:2005ew},  which we denote as $(\boldsymbol{\varphi},\boldsymbol \omega)$. 

\begin{table}[t!]
    \centering
    \begin{tabular}{|c|c|c|c|c|}
    \hline
     $S$ & $n_\text{f}$ & $n_-$ & zero mode error & $\lambda_1$ \\
     \hline
     $4$ & $81$ & $-3$ & 0.449 & 3.8009\\
     \hline
     $5$ & $121$ & $-3$ & 0.304 & 3.8028\\
     \hline
     $10$ & $441$ & $-3$ & 0.084 & 3.8055\\
     \hline
     $14$ & $841$ & $-3$ & 0.045 & 3.8060\\
     \hline
     $20$ & $1681$ & $-3$ & 0.022 & 3.8062\\
     \hline
     $25$ & $2601$ & $-3$ & 0.014 & 3.8063\\
     \hline
     $35$ & $5041$ & $-3$ & 0.0074 & 3.8064\\
     
     \hline  
    \end{tabular}
     \caption{The numerical results for the $Q = 0.2$ lukewarm solution using the fuzzy sphere method. }
    \label{table:fuzzydata}
\end{table}

We need a set of ``fuzzy rules" to implement our eigenvalue problem. The fuzzy Laplacian acts as nested commutators:
\begin{equation}\label{fuzzylap}
\nabla^2_{S^2}\varphi \quad \longrightarrow \quad -\sum_{i=1}^3[J_i,[J_i,\boldsymbol{\varphi}]]\,,
\end{equation}
while the fuzzy multiplication by coordinates on the sphere is implemented as an anti-commutator:
\begin{equation}\label{fuzzyprod}
x_i\varphi \quad \longrightarrow  \quad \frac{1}{2}\left(\frac{J_i}{\sqrt{S(S+1)}}\boldsymbol{\varphi}+\boldsymbol{\varphi}\frac{J_i}{\sqrt{S(S+1)}}\right)\,.
\end{equation}
With the rules (\ref{fuzzylap}) and (\ref{fuzzyprod}),
the differential operator $\mathcal{D}$ becomes a linear map on matrices, i.e., a superoperator. 

Practically, we can use the Choi-Jamiolkowski map to turn the superoperator into a linear map on  a doubled vector space:
\begin{equation}
\nabla_{S^2}^2\quad \longrightarrow \quad \nabla^2_\text{f}\equiv-\sum_{i=1}^{3}(J_i\otimes1-1\otimes J_i^\intercal)^2
\end{equation}
In our case, the differential operator $\mathcal D$ contains scalar multiplication with functions $U'',U',U,\Omega^{-2}$. Due to the $U(1)$ symmetry of the lukewarm background, these are functions of $\phi(\theta)$ only, or equivalently, function of $z\equiv\cos\theta$. So, any function of $z$ has a fuzzy version:
\begin{equation}
F(z) \quad \longrightarrow \quad  F_\text{f} = \frac{1}{2}\left[F\left(\frac{J_3}{\sqrt{S(S+1)}} \right)\otimes1+1\otimes F\left(  \frac{J_3}{\sqrt{S(S+1)}} \right)\right]\,.
\end{equation}
With these ingredients, the fuzzy version of the differential operator $\mathcal{D}$ is now an $2(2S+1)^2$-by-$2(2S+1)^2$ dimensional matrices
\begin{equation}  
\mathcal D_\text{f}\equiv\begin{pmatrix}
-U''_\text{f}/2 & \Omega^{-2}_\text{f}\nabla^2_\text{f}-U'_\text{f}\\
\Omega^{-2}_\text{f}\nabla^2_\text{f}-U'_\text{f} & -2U_\text{f}
\end{pmatrix},\ \ \ \mathcal D_\text{f}\Psi_\text{f}=\lambda\Psi_\text{f}\,.
\end{equation}

\noindent Under this method,  the (regularized) number of negative mode is defined as:
\begin{equation}
n_-=\text{number of negative eigenvalue}-n_\text{f},\quad \quad n_\text{f}\equiv(2S+1)^2
\end{equation}

For the specific example $Q=0.2$, which can be compared with  \cref{table:meshdata} for consistency, we report some results in  \cref{table:fuzzydata}.

From the numerically obtained spectrum, we indeed observe 5 nearly zero modes, which converges to exact zero modes in continuous limit (the `zero mode error' column in the above table record the largest absolute value of eigenvalue out of five nearly zero modes), and also $n_-=-3$ negative mode. We have also numerically verified this counting over the entire lukewarm branch $0<Q<1/4$. The first nonzero eigenvalue $\lambda_1$ (last column) converges, and is consistent with the mesh method, see  \cref{table:meshdata}.

A benefit of the fuzzy sphere method compared to the mesh method is that it preserves the background $U(1)$ symmetry. This is helpful if one would want to go to larger system size.

\subsection{Method III: Chebyshev-Gauss-Lobatto collocation}
\label{app: Chebyshev method detail}

As mentioned in \cref{sec:mesh}, the method that is actually the most efficient and precise is the Chebyshev-Gauss-Lobatto (CGL) collocation method~\cite{Trefethen2000Spectral}. This method utilizes the $U(1)$ symmetry of background metric, and directly works in sectors of fixed Matsubara quantum number $m$. In each sector, one solves the eigenvalue problem of a one-dimensional differential operator\footnote{See \eqref{soln2d} for definition of $A(\phi)$.}
\begin{equation}
\begin{bmatrix}
-A'''/2& A\partial_{\phi}^2+A'\partial_{\phi}-A''-A^{-1}(\frac{2\pi m}{\beta})^2\\
A\partial_{\phi}^2+A'\partial_{\phi}-A''-A^{-1}(\frac{2\pi m}{\beta})^2 & -2A'
\end{bmatrix}
\begin{bmatrix}
\varphi_m\\\omega_m
\end{bmatrix}=\lambda\begin{bmatrix}
\varphi_m\\\omega_m
\end{bmatrix}
\end{equation}
via discretizations on a Gauss-Lobatto lattice~\cite{Trefethen2000Spectral}. One truncates the continuum problem by expanding the functions into a basis of the first $n_\text{CGS}$ Chebyshev polynomials. 
The 2D smoothness condition requires that $\varphi_m,\omega_m$ go like $\sim(\phi-\phi_b)^{|m|/2}$, $\sim(\phi_c-\phi)^{|m|/2}$ when approaching the black hole and the cosmic horizon. Practically, one way to impose this boundary condition is to perform a field redefinition via $\varphi_m\equiv A^{|m|/2}\tilde{\varphi}_m,\  \omega_m\equiv A^{|m|/2}\tilde{\omega}_m$, and impose free boundary condition for $\tilde\varphi_m,\tilde\omega_m$.
We refer the readers to section 2.2.1 of~\cite{Kolanowski:2024zrq} for explicit detailed implementation of the Chebyshev-Gauss-Lobatto collocation method, and more applications in black hole spectrum. Our case is parallel to theirs.

\begin{table}[t]
  \centering
  \begin{minipage}[t]{0.48\linewidth}
    \centering
    \begin{tabular}{|c|c|c|}
    \hline
     $n_{\text{CGL}}$  & zero mode error & $ \lambda_1$  \\

     \hline
     $10$  & $1.4\times 10^{-2}$ & $3.8070$\\
     \hline
     $20$  & $2.4\times 10^{-7}$ & $3.80650423$\\

     \hline
     $50$  & $<10^{-16}$ & $3.80650437$\\

     \hline  
    \end{tabular}
  \end{minipage}\hfill
  \begin{minipage}[t]{0.48\linewidth}
    \centering
    \begin{tabular}{|c|c|c|}
    \hline
    $Q$ & $\lambda_1$ & $\lambda_{1, \,\textrm{pert}}$\\
    \hline
    $0.24$ & $2.9291$ & $7.68$ \\
    \hline
    $0.249$ & $0.6808$ & $0.768$ \\
    \hline
    $0.2499$ & $0.0759$ & $0.0768$ \\
    \hline
    $0.24999$ & $0.00767$ & $0.00768$ \\
    \hline
    \end{tabular}
  \end{minipage}

  \caption{\textbf{Left: }The numerical results for the $Q = 0.2$ lukewarm solution using the Chebyshev-Gauss-Lobatto collocation method. \textbf{Right: }The numerical results for the first non-zero eigenvalue in $m=0$ sector for various value of $Q$ via collocation method ($n_\text{CGL}=300$), compared with analytical perturbation results. }
  \label{table:meshdata chebyshev}
\end{table}

Now we may report our results.  We find one approximate zero mode in the $m = 0$ sector and two each in the $m = \pm 1$ sectors. As a cross check for previous two methods, at $Q=0.2$ the first non-zero mode appears in $m=0$ sector, with eigenvalue $3.8065$. This number is consistent with the other two methods. From \cref{table:meshdata chebyshev}~(left), we see that this method converges very fast even for relatively small $n_{\text{CGL}}$.

As another demonstration of the precision of the CGL method, we can use it to confirm that there is an additional zero mode as we approach $Q\rightarrow\frac{1}{4}$ from the lukewarm branch. Numerically, we find that this mode comes from the lowest non-zero mode in the $m=0$ sector, and becomes an exact zero mode in $\ell=1,m=0$ sector of the $Q=\frac{1}{4}$ charged Nariai black hole. In fact, one can  calculate analytically how does this zero mode get lifted in perturbation theory. Denote $Q=\frac{1}{4}-\varepsilon^2$,  one find that the lifted eigenvalue to be $\lambda_{1, \,\textrm{pert}} = 768\varepsilon^2+O(\varepsilon^3)$. We can compare the numerical value to the perturbative computation, and we find good match, see \cref{table:meshdata chebyshev}~(right).

\section{The analytic spectrum for charged Nariai}

The charged Nariai background has a round sphere metric and constant dilaton, and thus analytic computation of the spectrum of $\mathcal{D}$ becomes possible. In this Appendix, we will present such an analysis, for a general dilaton potential $U(\phi)$. The full higher dimensional problem is also analytically tractable \cite{Zimo}.

\label{app: nariai spectrum}
\subsection{Charged Nariai}

As discussed in the main text, whenever $U(\phi_0)=0$ and $U'(\phi_0)<0$, we can find a classical solution with $\phi =\phi_0$ and the background metric being a round sphere with radius $r_0=\sqrt{-2/U'(\phi_0)}$. Expanding around this saddle, the quadratic action for the fluctuations~\eqref{quad}, and the inner product become
\begin{equation}
I=\frac{1}{4G}\int_{S^2}(\varphi\nabla_{S^2}^2\omega+\omega\nabla_{S^2}^2\varphi)+4\varphi\omega+\frac{1}{2}b\varphi^2,\ \ b\equiv-r_0^2U''(\phi_0)
\end{equation}
\begin{equation}
\text{Inner Product}=\frac{1}{4G}\int_{S^2}r_0^2(\varphi^2+\omega^2)
\end{equation}
where $\nabla_{S^2}^2$ is the Laplacian on unit-radius round sphere. Note that the only dependence on the dilaton potential is through its first and second derivative at $\phi_0$.

Due to the rotation symmetry of the background, the spectrum can be easily solved in terms of the angular momentum quantum number $\ell$. The eigenvalues are given by
\begin{equation}
\lambda_\pm(\ell)=r_0^{-2}\left[\frac{b}{4}\pm\sqrt{\left(\frac{b}{4}\right)^2+\left(\ell(\ell+1)-2\right)^2}\right],\ \ \ell=0,1,2,3,...
\label{eq: Nariai spectrum}
\end{equation}
For $b\neq0$, there are three zero mode in $\ell=1$ sector, corresponding to the three CKVs (which are not Killing vectors) of round sphere. When continuously varying $b$ from negative to positive,  there are three negative modes in the $\ell=1$ sector crossing zero and become positive mode, causing $n_-$ to jump from $n_-=0$ to $n_-=-3$, as observed in numerics. In the marginal case $b=0$, there will six zero modes in $\ell=1$ sector in total, that contains five CKVs and one physical zero mode. So the one-loop partition function is divergent and the phase of partition function need to be determined at higher loop.

We now apply our general analysis to the specific  case of the charged Nariai branch of magnetic black hole, with $U(\phi)$ given in~\eqref{Uphi}. We find
\begin{equation}
\phi_0=\frac{1+\sqrt{1-12Q^2}}{6},\quad U'(\phi_0)=-\frac{12\sqrt{6}\sqrt{1-12Q^2}}{(1+\sqrt{1-12Q^2})^{3/2}}<0,
\end{equation}
and
\begin{equation}
    b=\frac{3(1-8Q^2-\sqrt{1-12Q^2})}{Q^2\sqrt{1-12Q^2}}\,.
\end{equation}
We find that $b$ has the same sign as $Q- \frac{1}{4}$. 
 Therefore, for charged Nariai with $Q < \frac{1}{4}$ we have $n_-=0$ and for charged Nariai with $Q > \frac{1}{4}$ we have $n_-=-3$. 
 
\begin{figure}[t!]
    \centering
    \includegraphics[width=1\linewidth]{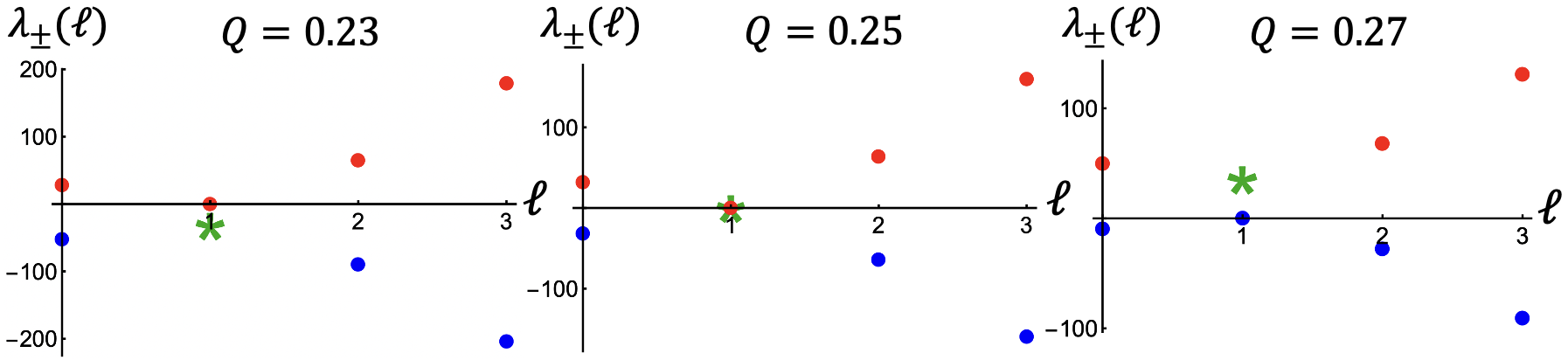}
    \caption{The spectrum of fluctuations (within the dilaton gravity reduction) for the charged Nariai background with various values of $Q$.}
    \label{fig: nariai spectrum}
\end{figure}

To help better visualize how modes in the $\ell = 1$ sector cross zero,  we plot the spectrum for $\ell \leq 3$ for various $Q$ in Figure~\ref{fig: nariai spectrum} (focus on the green markers).

\subsection{Ultracold Nariai limit}
One may wonder how the spectrum in~\eqref{eq: Nariai spectrum} approaches the spectrum of  $R^2$ gravity derived in the main text~\eqref{eq: ultracold spectrum} as we tune $Q$ towards the ultracold limit. Define $Q=\sqrt{1/12-3q^2}, c=36\sqrt{6}$ as in~\eqref{eq: ultracold scaling}, a local observer sitting along the equator will feel a temperature:
\begin{equation}
T_\text{loc}\equiv\frac{1}{2\pi r_0}=\frac{\sqrt{c}}{2\pi}q^{\frac{1}{2}}+O(q^{\frac{3}{2}})
\label{eq: ultracold temperature}
\end{equation}
As approaching ultracold limit,  $T_\text{loc}$ will go to zero as $q^{1/2}$. Fixing $\ell$, the spectrum becomes:
\begin{equation}
\lambda_+(\ell)=c+O(q),\ \ \lambda_-(\ell)=-cq^2[\ell(\ell+1)-2]^2+O(q^3)
\end{equation}
We see that $\lambda_+$ branch is the massive mode $\varphi$ we integrated out in~\eqref{eq: ultracold scaling}, and $\lambda_-$ matches the $R^2$ gravity's spectrum in~\eqref{eq: ultracold spectrum}.

\section{Unstable lukewarm solution}
\label{app: unstable lukewarm} 

In the main text, we focused on the lukewarm solution in dilaton gravity from dimensionally reducing 4D Einstein-Maxwell theory, which is always stable. For generic dilaton potential $U(\phi)$, one can find analog lukewarm solutions that are thermodynamically unstable. This depends on whether the total entropy is a local maximum or minimum. Since the cosmic horizon $\phi_c$ is related to the black hole horizon $\phi_b$ via equation $\int_{\phi_b}^{\phi_c} U=0$. We have the second derivative $\partial_{\phi_b^2}(\phi_b+\phi_c)={U'(\phi_c)-U'(\phi_b)\over U(\phi_b)}$ at the lukewarm geometry, so the stability condition is equivalent to the positivity condition for $U'(\phi_b)-U'(\phi_c)$.

We numerically computed the spectrum for  stable/unstable lukewarm solution in various dilaton potentials. We always find for the stable case: $n_{\text{zero}}=5$ and $n_-=-3, n_+=-2$; and for unstable case: $n_-=-2, n_+=-3$. In fact, if two dilaton potentials are related by a reflection $\tilde U(\phi)\equiv-U(-\phi)$. A stable lukewarm solution will be mapped to an unstable lukewarm solution. The spectrum of these two solutions will also be flipped, namely $n_{\pm}(\text{unstable})=n_{\mp}(\text{stable})$.

Apart from reflection, one can also go from a stable lukewarm to an unstable lukewarm by smoothly tuning dilaton potential to change $U'(\phi_b)-U'(\phi_c)$ from positive to negative. There will be one positive mode crossing zero and become negative mode, make $n_-$ jump from $-3$ to $-2$. Using collocation method in \cref{app: Chebyshev method detail} and resolve $U(1)$ symmetry of background metric, we find this mode is in zero Matsubara frequency sector.

\section{Observer model in dS$_4$}\label{app:highD}

In \cref{sec:dS3phase}, we discussed an observer model of a conical defect in dS$_3$ and a model integral for its state-counting partition function. In this Appendix, we discuss a similar setup in $D = 4$.

We consider a model for the observer as a perfect fluid in the center of the static patch, occupying a region of radius $r_0$. We work in units where $\Lambda = 3$ and consider the case where the fluid occupies a small region in de Sitter space, $r_0 \ll 1$. Therefore, we can approximate the background spacetime as flat space in the region of the fluid. Assuming spherical symmetry, one can write down an effective action describing the coupling between metric fluctuations and the fluid, valid to leading order in the perturbation:
\begin{equation}\label{fluidaction}
    I = - \frac{1}{16\pi G} \int \d^4 x\,\sqrt{g} (R- 6 ) +  \int d^4 x\, \sqrt{g_0} \left[ \left(1 +  \Phi (r) \right) \rho(r) -  \varepsilon(r) P(r)  \right].
\end{equation}
Here $\rho(r), P(r)$ are the density and the pressure of the fluid, and the perturbed metric is given by
\begin{equation}\label{metricfluid}
    ds^2 = \left(1 - r^2  + 2 \Phi (r) \right)d\tau^2  +\frac{dr^2}{1 - r^2  - 2\varepsilon(r) }  + r^2 d \Omega^2\,.  \quad \tau \sim \tau + \beta\,.
\end{equation}
In (\ref{fluidaction}), $g_0$ denotes the unperturbed metric. 
In the limit of small backreaction, we can ignore the contribution of the pressure term in (\ref{fluidaction}), effectively treating the fluid as dust. Under the flat space approximation, the equations for $\Phi$ and $\varepsilon$ are 
\begin{equation}
\begin{aligned}
     \frac{1}{r^2} \frac{d}{dr} (r^2  \Phi') &   = 4 \pi G  \rho, \\
     r  \Phi' & = \varepsilon\,.
\end{aligned}
\end{equation}
For a fluid with constant energy density $\rho = \rho_0$, we find solution
\begin{equation}\label{Phivarepsilon}
     \Phi = \frac{2\pi G r^2}{3} \rho_0 - 2 \pi G \rho_0 r_0^2\,, \quad \varepsilon = \frac{4\pi G r^2}{3} \rho_0\,, \quad r< r_0\,.
\end{equation}
Outside the fluid, $r>r_0$, we simply have the Schwarzschild solution
\begin{equation}
    \Phi = - \varepsilon = - \frac{G\mu}{r}, \quad \mu = \frac{4\pi }{3} \rho_0 r_0^3\,.
\end{equation}

Now, using the same idea as in  \cref{sec:dS3phase}, we would like to study the off-shell action of (\ref{metricfluid}), with $\rho_0$ (or equivalently the total mass $\mu$) and $\beta$ being simply parameters that are not necessarily at their on-shell values. By being off-shell, the parameter $\rho_0$ in the metric does not need to equal to the actual density of the fluid, which we denote as $\bar{\rho}$. We also denote the total mass of the fluid as $\Eo = \frac{4\pi}{3} \bar{\rho} r_0^3$. 

The action of the solution comes from several parts. First, we can evaluate (\ref{fluidaction}) in the flat space region, and get 
\begin{equation}\label{Iflat}
    I_{\textrm{flat}} \approx - \beta\frac{ \mu}{2} + \beta \frac{3G}{5 r_0 } \mu^2 + \beta \left( 1-  \frac{6 G}{5 r_0 }\mu \right)  \Eo \,.
\end{equation}
We then have the bulk action from outside the fluid to the cosmic horizon $r = r_h$. Since $r_0\ll 1$, we can approximate the action as simply from $0$ to $r_h$, given by 
\begin{equation}\label{Ibulk}
    I_{\textrm{bulk}} = -\frac{1}{16\pi G} \int d^4 x \sqrt{g} (12- 6) \approx  -\frac{\beta}{4 G} \int_{0}^{r_h} dr 6 r^2  = - \frac{\beta}{2 G} r_h^3,
\end{equation}
where $r_h$ is implicitly a function of $\mu$. Finally, we have the contribution from the cosmic horizon, due to the existence of a conical defect
\begin{equation}\label{Ihor}
    I_{\textrm{hor}} = - \frac{1}{16\pi G} 4 \pi r_h^2 (4\pi + \beta f'(r_h)), \quad f = 1 - r^2- \frac{2G\mu}{r}\,. 
\end{equation}
Combining (\ref{Iflat}) $\sim$ (\ref{Ibulk}) and expanding at small $\mu$, we find 
\begin{equation}
    I \approx  -\frac{\pi}{G} + (2\pi - \beta) \mu + \beta \frac{3G}{5r_0} \mu^2 +2\pi G \mu^2 + \beta \left( 1-  \frac{6 G}{5 r_0 }\mu \right)  \Eo\,.
\end{equation}
The saddle point for $\beta$ and $\mu$ is at
\begin{equation}
    \beta \approx 2\pi + 4 \pi G \Eo , \quad \mu = \Eo - \frac{3 G}{5 r_0 } \Eo^2\,. 
\end{equation}
Expanding $\beta$ and $\mu$ around the saddle point, we find
\begin{equation}\label{int4d}
    I \approx - \frac{\pi}{G} + 2\pi \Eo - \frac{6\pi G }{5 r_0 } \Eo^2 - \delta\beta \delta\mu + \frac{6\pi G }{5 r_0 } \delta\mu^2 \,.
\end{equation}
This expression can be compared with (\ref{Z3d}) in the $D=3$  case. Here, the extra ingredient is that there is an additional $\delta \mu^2$ term. 

In the prescription of \cite{Maldacena:2024spf}, an important ingredient that would affect the sign of the final answer is whether the final integral over the length of the observer worldline is stable. There, it was claimed that the integral is stable. We can verify this using our (\ref{int4d}). First, we have to recognize that the $\beta$ in (\ref{metricfluid}) is not the length of the circle felt by the observer $\beta_{\textrm{obs}}$. In our case, since the fluid is smeared out, there could be many different notions of $\beta_{\textrm{obs}}$. One possible definition is to take $\beta_{\textrm{obs}}$ as the size of the circle at the center of the fluid, which will mean
\begin{equation}
    \beta_{\textrm{obs}} \approx  (1 + \Phi (0)  ) \beta = \left(1 - \frac{3}{2} \frac{G\mu}{ r_0 } \right) \beta\,.
\end{equation}
Therefore, we have
\begin{equation}
    \delta \beta_{\textrm{obs}} \approx \delta \beta - \frac{3 \pi G}{ r_0 }  \delta \mu \,.
\end{equation}
Using $\delta \beta_{\textrm{obs}}$ and $\delta\mu$, the relevant terms in (\ref{int4d}) becomes
\begin{equation}\label{intoutmu}
     I \supset - \delta\beta_{\textrm{obs}} \delta\mu - \frac{9 \pi G}{5 r_0 } \delta\mu^2\,.
\end{equation}
Integrating out $\delta\mu$, one indeed finds that the $\delta \beta_{\textrm{obs}}$ mode is stable, agreeing with the claim in \cite{Maldacena:2024spf}. One could attempt other reasonable definitions of $\delta\beta_{\textrm{obs}}$, such as by averaging the length of the $\tau$ circle over the fluid, but the conclusion remains unchanged.

Of course, in the prescription we were advocating in  \cref{sec:statecount}, whether the $\delta\beta_{\textrm{obs}}$ mode is stable or unstable does not play an important role.

\bibliography{references}

\bibliographystyle{utphys}

\end{document}